\documentclass[letterpaper,twocolumn,10pt]{article}
\usepackage{usenix}

\usepackage{graphicx}
\usepackage{amsmath}
\usepackage{xspace}
\usepackage{booktabs}
\usepackage{float}
\usepackage{enumitem}
\setlist[enumerate]{nosep, topsep=2pt, leftmargin=1.5em}
\usepackage{colortbl}
\definecolor{groupbg}{gray}{0.92}
\usepackage{xurl}

\usepackage{tcolorbox}
\tcbuselibrary{breakable, skins}
\usepackage{soul}
\usepackage{multirow}
\usepackage[capitalize,noabbrev]{cleveref}

\definecolor[named]{ACMPurple}{cmyk}{0.55,1,0,0.15}
\definecolor[named]{ACMDarkBlue}{cmyk}{1,0.58,0,0.21}
\hypersetup{colorlinks,
  linkcolor=ACMDarkBlue,
  citecolor=ACMPurple,
  urlcolor=ACMDarkBlue,
  filecolor=ACMDarkBlue}

\definecolor{reasonMemBg}{HTML}{E6EDF3}
\definecolor{reasonObjBg}{HTML}{F4E7DD}
\definecolor{reasonKnoBg}{HTML}{E7F0EA}

\newcommand{\reasonM}[1]{{\sethlcolor{reasonMemBg}\hl{\textit{#1}}}}
\newcommand{\reasonO}[1]{{\sethlcolor{reasonObjBg}\hl{\textit{#1}}}}
\newcommand{\reasonK}[1]{{\sethlcolor{reasonKnoBg}\hl{\textit{#1}}}}

\DeclareRobustCommand{\cnum}[1]{\mbox{\textcircled{\scriptsize #1}}\,}

\newcommand{\etal}{et~al.\xspace}

\newcommand{\securevibebench}{{\sc SecureVibeBench}\xspace}
\newcommand{\susvibes}{{\sc SusVibes}\xspace}
\newcommand{\ase}{{\sc A.S.E}\xspace}
\newcommand{\asleep}{{\sc Asleep}\xspace}
\newcommand{\secodeplt}{{\sc SeCodePLT}\xspace}
\newcommand{\vibeapps}{{\sc VibeApps}\xspace}
\newcommand{\vibevulns}{{\sc VibeVulns}\xspace}

\definecolor{findingaccent}{HTML}{374151}
\definecolor{findingbg}{HTML}{F3F4F6}
\newtcolorbox{findingbox}[1][Takeaway]{
  enhanced, breakable,
  colback=findingbg, frame hidden,
  boxrule=0pt,
  arc=6pt, rounded corners,
  left=5pt, right=5pt, top=5pt, bottom=5pt,
  before skip=5pt, after skip=0pt,
  fontupper=\small,
  detach title,
  before upper={\textcolor{findingaccent}{\textbf{\small #1.}}\hspace{0.6em}\ignorespaces},
}

\begin{document}

\date{}

\title{\Large \bf Understanding the (In)Security of Vibe-Coded Applications}

\author{
{\rm Junquan Deng}\\
Independent Researcher\\
\texttt{junquanpdeng@gmail.com}
\and
{\rm Zhiyu Fan}\\
Microsoft\\
\texttt{fanzhiyu@microsoft.com}
\and
{\rm Ruijie Meng}\\
National University of Singapore\\
\texttt{ruijie\_meng@u.nus.edu}
}

\maketitle

\begin{abstract}
Recent advances in large language models (LLMs) have enabled vibe coding, an emerging software development paradigm in which users create applications primarily through natural-language interactions with AI agents. Due to its low barrier to entry, vibe coding is rapidly gaining adoption in practice. Unlike conventional AI-assisted programming, where developers remain responsible for implementation and code review, vibe coding delegates a substantial portion of the development process to AI systems. This shift raises a fundamental question: how (in)secure are applications developed through vibe coding? In this paper, we conduct a systematic study of the security of real-world vibe-coded applications. We collect 9,041 open-source applications developed using popular AI agents (Claude Code and Lovable), and audit 200 publicly deployed applications, uncovering 1,186 vulnerabilities. Our study of these applications and vulnerabilities reveals several key findings: (1) insecurity is the norm rather than the exception: 91.0\% of audited applications contain at least one vulnerability, and 65.77\% of identified vulnerabilities are rated Critical or High severity, concentrated in broken access control, injection, and authentication failures; (2) these vulnerabilities are traceable to eight recurring failure modes rooted in three systematic limitations of AI agents: memory defects, objective defects, and knowledge defects; and (3) while improved agent harness and prompting strategies can reduce the incidence of vulnerabilities, they do not eliminate the underlying security risks. Overall, our study provides an empirical understanding of the security landscape of vibe-coded applications and lays the groundwork for addressing the security risks in the growing delegation of software development to AI systems.
\end{abstract}

\section{Introduction}
\label{sec:introduction}

Recent advances in large language models (LLMs) are fundamentally reshaping how software is developed~\cite{se3.0}. What began as code-completion assistants~\cite{copilot, codex} capable of generating individual functions or features has rapidly evolved into autonomous AI coding agents~\cite{lovable, claudeCode} that design, implement, and deploy complete, runnable software applications from natural language instructions. This evolution has led to a new development paradigm commonly referred to as \textit{vibe coding}~\cite{karpathy2025vibe, zhao2025vibecoding}. In this paradigm, users describe high-level requirements in natural language and rely on AI coding agents to generate entire software applications from scratch, with minimal human intervention. 

Since its introduction, vibe coding has rapidly gained traction across the software development ecosystem, evolving from an experimental concept into a
production reality. The appeal of vibe coding lies in substantially lowering the technical barrier to application creation by abstracting away much of the development process. As a result, individuals with limited programming experience can now create and deploy applications that previously required significant development expertise. AI agents such as Claude Code~\cite{claudeCode} and Lovable~\cite{lovable} have attracted millions of users and enabled the creation of applications at an unprecedented scale. For example, Lovable reported more than eight million users and over 100,000 new applications created every day as of early 2026~\cite{lovableStatistics}. Moreover, more than 10\% of the applications are deployed as live, publicly accessible systems, as we later show in this study.

With the popularity of vibe coding, a critical concern arises: \emph{how (in)secure are vibe-coded applications in practice?} In the vibe-coding paradigm, even non-expert individuals can create and directly deploy applications without fully understanding the underlying implementation, which may leave security-critical decisions made by AI agents unaudited. This concern is not merely hypothetical, and practitioners continue to report applications that expose serious vulnerabilities~\cite{databricks2025passing,owasp2025top10,vibegraveyard2025}. For example, an official vulnerability record (CVE-2025-48757, CVSS 9.3) documented insufficient Row-Level Security policies in the databases backing sites generated by Lovable, allowing unauthenticated attackers to access arbitrary user data~\cite{nvd2025cve48757}. Similarly, in our study, we identified a publicly deployed vibe-coded application: a healthcare provider platform that exposed a large volume of customer information due to broken access control. We also identified a command injection vulnerability that enabled remote code execution and full control of its backend. Such incidents underscore the need for a systematic study of the security of vibe-coded applications.

Although a growing body of work has examined the security of LLM-generated code~\cite{pearce2022asleep, lian2025ase, zhao2025vibecoding, chen2025securevibebench}, prior studies fall short of capturing the security risks of vibe-coded applications. Existing research has primarily focused on evaluating the security of AI-assisted programming. In this paradigm, AI coding agents operate within established codebases and are tasked with completing specific development activities, such as implementing new features~\cite{zhao2025vibecoding, chen2025securevibebench}, where software development is largely reduced to mere code-filling tasks. This paradigm fundamentally differs from vibe coding, where AI coding agents are instructed to create entire software applications from scratch. In vibe-coded applications, security vulnerabilities may arise not only from insecure code generation, but also from architectural and trust-boundary decisions that AI agents make autonomously throughout the development lifecycle without much human review. Therefore, findings from prior studies cannot be generalized to characterize the security of vibe-coded applications.

\paragraph{Our Work.} To this end, we conduct the first systematic study to understand the security of vibe-coded applications. We construct \vibeapps, a large-scale corpus of 9,041 open-source vibe-coded applications. From this corpus, we randomly sample 200 deployed applications that are publicly accessible and use them as the target of our
security analysis. We then develop a human-in-the-loop vulnerability-discovery workflow, combining multi-agent code auditing with structured human validation. Using this workflow, we identify 1,186 vulnerabilities across the sampled applications, forming our vulnerability dataset \vibevulns. To the best of our knowledge, this is the first large-scale dataset for studying the security of real-world vibe-coded applications. In the study, we seek to answer the following research questions:
\begin{itemize}[leftmargin=*]
\item \textbf{RQ.1: What is the landscape of real-world vibe-coded applications?} We characterize the ecosystem of vibe-coded applications, including their scales, development characteristics, application categories, and deployment status.

\item \textbf{RQ.2: How prevalent and severe are vulnerabilities in deployed vibe-coded applications?} We measure the prevalence and severity of their vulnerabilities, including the distribution, to assess their overall security posture.

\item \textbf{RQ.3: What factors contribute to these vulnerabilities?} We examine how these vulnerabilities are introduced in the vibe-coding lifecycle.

\item \textbf{RQ.4: What conditions can mitigate the vulnerabilities of vibe-coded applications?} We examine whether advanced agent harnesses, stronger foundation models, and carefully-crafted user prompts can mitigate vulnerabilities in vibe-coded applications.

\end{itemize}

Through answering these questions, we characterize the security risks of vibe-coded applications and provide actionable insights for different stakeholders (e.g., agent and harness developers, and end users). Our key findings are summarized as follows. For RQ.1, we find that vibe coding is often used to develop large-scale applications within a short time frame. Moreover, web applications form the dominant application
category, and more than 11\% of these applications are publicly deployed. For RQ.2, we find that 91.0\% of deployed vibe-coded applications contain at least one vulnerability, with 65.77\% of identified vulnerabilities rated Critical or High severity and risk concentrated in broken access control, injection, and authentication failures. For RQ.3, we show that these vulnerabilities are traceable to eight recurring failure modes across three defect classes---memory, objective, and knowledge defects---that reflect systematic limitations of AI agents throughout the vibe-coding lifecycle. For RQ.4, nearly all investigated conditions mitigate security vulnerabilities to varying degrees, where production-ready prompting strategy and a security-hardened agent harness are the most effective. Nevertheless, mitigating vulnerabilities introduced by different factors requires different mitigation strategies.  

\paragraph{Contributions.} We make the following contributions:
\begin{itemize}[leftmargin=*]

\item We construct \vibeapps, a large-scale dataset of 9,041 real-world vibe-coded applications, and build \vibevulns, a dataset of 1,186 validated vulnerabilities from 200 publicly deployed vibe-coded applications. Both datasets serve as a foundation for future research in this domain. 

\item We present the first large-scale empirical study of the security of vibe-coded applications, providing a systematic characterization of the
prevalence, severity, and failure modes of security risks in vibe coding.

\item Our findings provide actionable guidance for securing vibe-coded applications, which requires safeguards embedded throughout the vibe-coding lifecycle beyond improvements to code-generation quality alone.  

\end{itemize}
\section{Background and Related Work}
\label{sec:background}

\paragraph{AI-Assisted Programming and Vibe Coding.} \label{sec:definition}

LLMs are increasingly used in software development through tools that support code completion, bug fixing, and even repository-level modification. We refer to this paradigm as \emph{AI-assisted programming}: LLMs
provide code-level assistance, while developers remain responsible for application design, implementation decisions, and code review. Industry-grade tools such as GitHub Copilot are now used by millions of developers~\cite{shani2023github}, leading to a substantial fraction of code in production repositories being generated with assistance from LLMs~\cite{chandra2024aise}.

Building on these advances, \emph{vibe coding} has emerged as an AI-led development paradigm in which users describe desired application behavior in natural language and rely on coding agents or development platforms to design, implement, and iteratively refine complete applications~\cite{karpathy2025vibe, chou2025building}.
Unlike AI-assisted programming, where humans remain the
primary authors and decision-makers, vibe coding shifts substantial authorship and oversight toward AI agents, leaving humans primarily specifying high-level goals and providing feedback. This distinction is central to our study because security-relevant decisions in vibe-coded applications may be made as part of an AI-led development workflow rather than through explicit human design and review.

\paragraph{Security of AI-Assisted Programming.}

In recent years, the security of AI-generated code has attracted growing attention. However, existing studies mainly focus on \emph{AI-assisted programming}, where LLMs generate code snippets or modifications integrated into existing codebases. Early work, including \asleep~\cite{pearce2022asleep} and \secodeplt~\cite{secodeplt}, evaluated the security of AI-generated code on synthetic CWE-based tasks involving individual functions or single-file programs. More recent studies, such as \ase~\cite{lian2025ase}, \securevibebench~\cite{chen2025securevibebench}, and \susvibes~\cite{zhao2025vibecoding}, have expanded the scope to repository-level settings, where coding agents modify multiple files within established codebases. Despite this increased realism, the common characteristic of these studies is that AI-generated code is evaluated in the context of existing software systems.

Evidence regarding the security of AI assistance is mixed. A large body of work has shown that LLMs frequently generate vulnerable code. For example, Pearce~\etal~\cite{pearce2022asleep} found that approximately 40\% of GitHub Copilot completions for CWE-targeted prompts contained security vulnerabilities. Subsequent studies have reported similar findings across a variety of models, programming languages, and task settings~\cite{khoury2023secure,tihanyi2025secure,fu2025security,schreiber2025security,shahid2025llmcsec,dora2025hidden,schaad2025still}. However, the presence of vulnerabilities in generated code does not necessarily imply that AI-assisted development produces less secure software applications, since developers may identify and correct these issues during review~\cite{perry2023users}.
Collectively, these findings suggest that LLMs often generate insecure code and that the ultimate security outcome depends critically on the expertise, vigilance, and review practices of the human developers.

\paragraph{Security of Vibe Coding.}

Despite this growing body of literature, to the best of our knowledge, no prior study has examined the security of real-world vibe-coded applications. Existing research focused on AI-generated code integrated into human-authored codebases~\cite{fu2025security,schreiber2025security} or benchmark tasks conducted in controlled experimental settings~\cite{lian2025ase,vero2025baxbench,zhao2025vibecoding,chen2025securevibebench}. Consequently, the current evidence is largely restricted to localized vulnerabilities evaluated against a predefined oracle, which cannot capture how application security emerges from the interplay of architectural decisions, implementation choices, and deployment configurations across an entire software system exposed to real users. Yet these higher-level decisions are precisely where autonomous coding agents exercise the greatest degree of control in the vibe-coding paradigm.

This gap is becoming increasingly consequential as vibe coding enters the mainstream. AI coding agents such as Claude Code and Lovable already enable large numbers of users, many with limited software engineering expertise, to build and deploy production applications. Security risks in these systems therefore have immediate implications for real users. To address this gap, we provide the first empirical characterization of the security risks of real-world vibe-coded applications. We move beyond controlled benchmarks to examine vulnerabilities emerging in real-world applications already operating on the public Internet.

\section{Dataset Construction}
\label{sec:setting}

To investigate the security risks of vibe-coded applications, we need a high-quality dataset of their vulnerabilities. To this end, we first construct \vibeapps, a representative corpus of real-world vibe-coded applications. We then develop a framework that combines multi-agent code auditing with human validation to discover vulnerabilities. The resulting vulnerabilities constitute our dataset \vibevulns, which serves as the foundation for the security analysis. 

\subsection{Vibe-Coded Application Collection}
\label{sec:method:dataset}

To collect real-world vibe-coded applications, the primary challenge lies in how to identify them, as they are not explicitly labeled as vibe-coded. To address this, we design a three-step pipeline. We first discover potential vibe-coded applications using fingerprints associated with AI coding agents. We then filter these candidates according to a set of criteria to obtain a high-quality subset. Finally, we determine applications to be vibe-coded based on the degree of AI contributions.  

\paragraph{Step 1: Potential Application Discovery.}

Although there are no explicit indicators for vibe-coded applications, the repositories with AI-generated code usually retain fingerprints left by the corresponding coding agents. For example, repositories using Claude Code include the configuration directory \texttt{.claude/}, while those developed with Lovable contain metadata \texttt{<meta name="author" content="Lovable" />}. These fingerprints provide useful signals for discovering potential vibe-coded applications. Based on this observation, we query the GitHub Code Search API to collect repositories that contain the relevant fingerprints. In this study, we focus on two popular AI coding agents: Claude Code, representing command-line-based agents, and Lovable, representing AI-based development platforms. They capture distinct development workflows and provide broader coverage of the current vibe-coding ecosystems.
This step results in an initial dataset of 74,800 repositories.

\paragraph{Step 2: High-Quality Application Selection.}

To ensure that our study is conducted on high-quality applications, we further select those from the initial dataset that are likely to be used in practice. Specifically, we select the repositories that satisfy the following criteria: (1) they provide sufficiently complete documentation, with a README containing at least 100 characters; (2) they have a GitHub repository size between 10 and 500{,}000 KB, 500 to 1M lines of code, at least five files, and 10 to 5{,}000 commits, where the lower bounds filter out small repositories with trivial functionality and the upper bounds filter out the extremely large repositories that typically contain substantial redundant implementations (e.g., third-party libraries) based on our observation; and (3) they are implemented as applications or tools, thereby excluding repositories intended primarily for documentation, tutorials or illustrative examples. After applying these filters, 38,078 repositories were retained from the initial dataset.

\paragraph{Step 3: Vibe-Coded Application Determination.} 

After the previous two steps, we obtain a high-quality dataset of real-world repositories that contain AI-generated code. We next determine whether an application is vibe-coded. Specifically, a vibe-coded application should be created by AI from scratch, instead of filling features in existing codebases. To capture this property, we require the initial commit of a repository to be AI-authored. In addition, vibe-coded applications should be developed predominantly by AI coding agents, with minimal human intervention. This requires us to identify AI-authored commits within each repository. Many coding agents and vibe-coding platforms leave their identification in commit messages, author fields, or commit trailers, such as \texttt{"Co-authored-by: Claude <noreply@anthropic.com>"} or \texttt{"Generated with Claude Code"}. Therefore, we label commits containing such markers as AI-authored. To quantify the degree of AI involvement, we measure two metrics: the proportion of AI-authored commits and the proportion of AI-authored code lines. We classify an application as vibe-coded only if both metrics are at least 90\%.
We selected 90\% because, in our inspection, repositories below this threshold still contained non-trivial human-authored logic, while a higher cutoff would exclude repositories whose human contribution was purely ancillary (e.g., README edits and dependency bumps). 
Applying these AI-authorship criteria retained 9,041 applications, forming our dataset \vibeapps.

\subsection{Vulnerability Discovery}
\label{sec:method:security}

After collecting vibe-coded applications, we further build their vulnerability dataset for security analysis. However, discovering vulnerabilities across these applications poses another challenge: they vary substantially in \emph{scale}, \emph{programming language}, \emph{technology stack}, and even \emph{application type} (as shown in \Cref{sec:rq1}). Conventional vulnerability-detection tools can hardly detect such a diverse set of applications effectively. To address the challenges, we adopt code-auditing agents for vulnerability discovery. Nevertheless, auditing different \emph{types} of applications requires different agent capabilities and auditing workflows. Therefore, we focus on the dominant type of vibe-coded applications: web applications. As shown in \Cref{sec:rq1}, among the 9,041 collected applications, 8,695 (over 96\%) are web applications, making their vulnerabilities broadly representative of the overall security landscape of vibe coding.

Based on this, we develop the vulnerability-discovery framework that combines multi-agent code auditing with structured human validation. Specifically, we first leverage a multi-agent system to analyze repositories and generate structured vulnerability reports. We then perform both AI-assisted and manual validation to select true vulnerabilities, producing the final set of vulnerabilities for security analysis.      

\paragraph{Step 1: Multi-Agent Code Auditing.}

We leverage LLM-based code auditors to identify potential vulnerabilities in repositories. To maximize vulnerability coverage, we adopt a $2\times2$ auditing design that combines two agent frameworks with two security auditing pipelines, yielding four independent audit reports for each repository. The two agent frameworks are Claude Code with Claude Sonnet 4.6 and GitHub Copilot with GPT-5.3-Codex. Each agent is paired with two complementary security skill sets: OpenAI's official security-best-practices skill~\cite{openaiSecuritySkill}, which provides guidance for detecting web application vulnerabilities across a range of technology stacks, and Antigravity Awesome Skills~\cite{antigravitySkills}, a widely adopted collection of security skills with nearly 40k GitHub stars that integrates conventional static application security testing tools and vulnerability-detection rules. 

Each skill set is organized around a primary security-analysis agent and a collection of technology-specific subagents. The primary agent, defined by the top-level \texttt{SKILL.md}, first analyzes the repository structure, framework usage, and technology stack. Based on this assessment, it automatically invokes specialized subagents tailored to the detected technologies. These subagents implement security auditing strategies specific to individual frameworks and ecosystems, such as \texttt{React} and \texttt{Node.js}. This hierarchical architecture helps reduce missed vulnerabilities arising from the limitations of any single model or auditing strategy. 

\begin{figure*}[htbp]
\centering
\setlength{\abovecaptionskip}{1pt}
\setlength{\belowcaptionskip}{-5pt}
\includegraphics[width=1.0\textwidth]{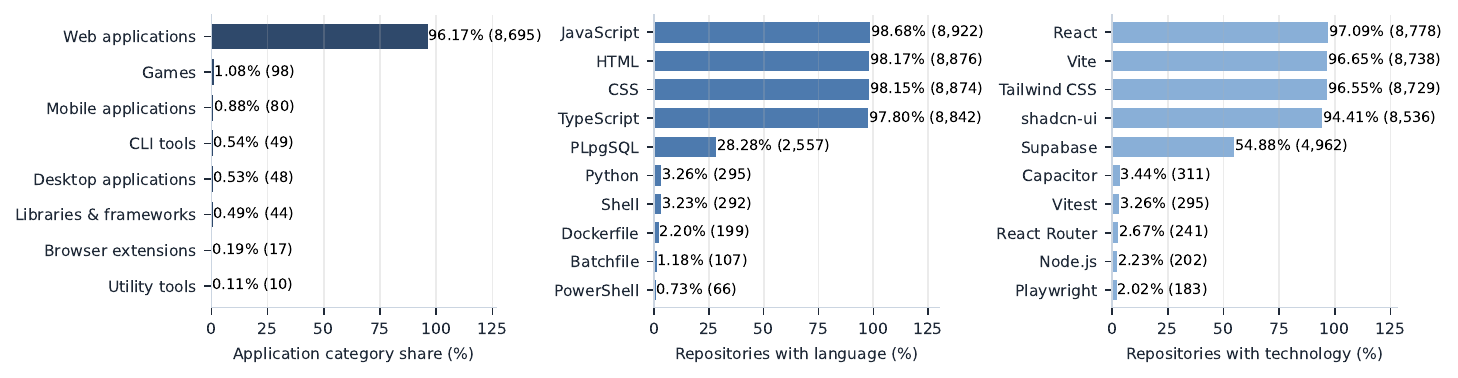}
\caption{Application categories and the most-used development languages and technology stacks in vibe-coded applications.}
\label{fig:rq1-tech}
\end{figure*}

We apply this $2\times2$ design to audit web applications. To ensure that the discovered vulnerabilities reflect realistic security exposure, we restrict our analysis to applications with publicly accessible deployments. These account for 984 of the 8,695 web applications (11.32\%, as shown in \Cref{sec:rq1}). Auditing all deployed applications would require substantial computational resources and manual validation effort.
Therefore, we randomly sampled 200 applications from the 984 deployed repositories for detailed security auditing. The whole audit process consumed 2.85 billion tokens and 128.47 hours of compute time. It yielded 800 audit reports and contained 11,150 raw vulnerability reports.

\paragraph{Step 2: Vulnerability Deduplication and Validation.}

The raw vulnerability reports contain substantial redundancy, as the four auditors frequently identify the same vulnerability using different descriptions. To address this, we review the four audit reports for each repository with assistance from AI agents. Specifically, we normalize their outputs, deduplicate overlapping candidate vulnerabilities, and merge semantically similar reports into canonical entries. We further examine the vulnerability-introducing commits to ensure that all retained vulnerabilities were introduced by AI. As a result, we obtain 2,125 vulnerability candidates.

To assess whether these candidates are true positives, we first built a validation agent that investigates the exploitability of each reported issue. The agent examines the relevant source files, traces whether user-controlled input can reach the reported sink, and checks for any sanitization or guard conditions along the execution path. It then attempts to construct a harmless proof-of-concept probe to the deployed application that would trigger the vulnerability if it were exploitable. Vulnerability candidates with clearly invalid exploitation paths are discarded, while confirmed cases are forwarded for human review. This process retained 1{,}287 vulnerability candidates.

Each remaining candidate is then independently reviewed by two authors with security expertise. They reviewed its audit report, inspected the relevant application source code, and examined the deployed website to verify the exploitability. A candidate is only considered a vulnerability if both authors classify it as exploitable. Furthermore, the two authors assign each vulnerability a severity level (Low, Medium, High, or Critical) following OWASP risk-rating criteria~\cite{owaspRiskRatingMethodology}, and map it to the corresponding OWASP Top~10 category~\cite{owasp2025webtop10}. Disagreements are resolved through discussion. Inter-rater agreement was Cohen's $\kappa = 0.87$. This produced 1{,}186 vulnerabilities, which form our dataset \vibevulns. 

In addition, we assess whether our framework potentially misses vulnerabilities in two ways. First, we conduct a blind independent audit on a stratified random sample of 20 repositories from our corpus. Two security researchers uninvolved in framework design perform security reviews without access to our reports. The vulnerabilities they identified were already captured by our framework, corresponding to an estimated miss rate below 15\% given the sample size (95\% CI, Rule of Three). Second, we evaluate our framework on Juice Shop~\cite{juiceShop} with a continuously maintained set of known vulnerabilities, including some disclosed after our models' training cutoff. Our framework achieves a recall of 95.70\% on this dataset. Together, these results suggest that our framework does not systematically miss vulnerabilities.
\section{RQ.1: What Is the Landscape of Real-World Vibe-Coded Applications?}
\label{sec:rq1}

To answer RQ.1, we characterize the 9,041 applications in our corpus \vibeapps along four dimensions: repository scale, development characteristics, application categories, and deployment status. This characterization establishes the landscape of real-world vibe-coded applications and
motivates our subsequent security analysis of deployed web applications.

\paragraph{Repository Scale and Development Characteristics.}

Real-world vibe-coded applications are often large-scale. The median repository contains 8{,}254 lines of code (LoC) across 100 files, while repositories at the 90th percentile reach 20{,}239 LoC and 193 files. Overall, 95.62\% of repositories contain more than 5{,}000 LoC, and 34.32\% exceed 10{,}000 LoC. Despite their considerable scale, these applications are typically developed within a short time. Measured from the first to the last commit, the median development span is 8.39 days. Furthermore, 67.81\% of repositories have development spans no longer than 30 days, and 84.87\% are completed within 90 days. However, these short development spans do not imply limited development activities: 64.22\% of repositories contain at least 25 commits, indicating substantial development activity within compressed periods.

To better characterize the development process, we examine the relationships between repository scale, development span, and commit counts. Commit count correlates strongly with repository size (Spearman $\rho=0.60$ for LoC) but only moderately with development span ($\rho=0.40$). Accordingly, commit density drops sharply as development span increases, from a median of 316.41 commits per day for repositories completed within a day to 0.58 commits per day for those spanning over 90 days. This indicates that vibe-coded applications are typically built in short, high-intensity periods.

Taken together, these findings reveal a distinctive scale-and-speed profile: many vibe-coded applications combine large codebases with highly concentrated development activity. Rather than emerging through the gradual accumulation of small code generations, they are often produced through intense bursts of project-level iteration.
This development pattern suggests that human reviews and hardening are unlikely to be applied before deployment. This confirms our motivation to study the security of vibe-coded applications.

\paragraph{Application Categories and Deployment Status.}

\begin{figure}[t]
\centering
\setlength{\abovecaptionskip}{0pt}
\setlength{\belowcaptionskip}{-5pt}
\includegraphics[width=\columnwidth]{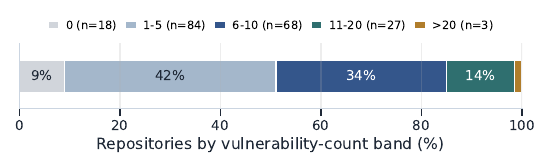}
\caption{Distribution of applications by vulnerability count.}
\label{fig:rq2-prevalence}
\end{figure}

We present the categories of vibe-coded applications, and their most-widely used programming languages and technology stacks in~\Cref{fig:rq1-tech}. It is interesting to note that vibe-coded applications are highly biased towards web applications, which account for 96.17\%. In contrast, the second-largest category, Games, only represents 1.08\% of the applications. The distributions of development languages and technology stacks further reinforce this trend. \texttt{JavaScript}, \texttt{HTML}, \texttt{CSS}, and \texttt{TypeScript} are the most commonly used programming languages, while \texttt{React}, \texttt{Vite}, \texttt{Tailwind CSS}, and \texttt{shadcn-ui} are the predominant technologies adopted in vibe-coded applications. This finding suggests that vibe coding is primarily used to develop web applications.
Applications developed by a single agent exhibit highly similar patterns.

We further examine whether the vibe-coded applications are actively deployed. Among all 9{,}041 applications, 1{,}007 (11.14\%) contain validated and reachable deployment links. 
Restricting our analysis to web applications, 984 of 8,695 (11.32\%) are actively deployed.
Therefore, vibe-coded applications are not merely repository artifacts, but also deployed systems with externally observable attack surfaces.

\begin{findingbox}[RQ.1 Findings]
Vibe-coded applications are typically both code-heavy and rapidly developed, indicating rapid project-level iteration rather than only small-scale
code generation. Moreover, they are strongly concentrated in web applications, accounting for about 96\%, and about 11\% are publicly deployed.
\end{findingbox}

\section{RQ.2: How Prevalent and Severe are Vulnerabilities in Vibe-Coded Applications?}
\label{sec:rq2}

To answer RQ.2, we analyze the 1{,}186 vulnerabilities from our dataset \vibevulns from two perspectives: how widely they appear across applications, and how severe they are in terms of OWASP category, a widely recognized taxonomy of web application security~\cite{owasp2025webtop10}.

\paragraph{Vulnerability Prevalence.} \Cref{fig:rq2-prevalence} shows that vulnerabilities are broadly distributed across vibe-coded applications. Of the 200 audited applications, 182 (91.00\%) contain at least one vulnerability. Among these vulnerable applications, the median number of vulnerabilities is 6, and the mean number is 6.52, with an interquartile range of approximately 3--9 and a 90th percentile of 12. These results indicate that vulnerabilities are highly prevalent across vibe-coded applications.

\begin{table}[t]
\centering
\setlength{\abovecaptionskip}{5pt}
\setlength{\belowcaptionskip}{0pt}
\caption{Vulnerability count and density by repository size.}
\label{tab:size-groups}
\small
\setlength{\tabcolsep}{3pt}
\begin{tabular}{lcrrr}
\toprule
Size & \#LoC & \#Repos & \shortstack{\#Vuln/repo} & \shortstack{VulnDensity/kLoC} \\
\midrule
Small & $\leq$5k & 15 & 2.13 & 0.85 \\
Medium & 5k--8k & 78 & 4.31 & 0.67 \\
Large & 8k--20k & 83 & 6.28 & 0.53 \\
xLarge & $>$20k & 24 & 12.38 & 0.22 \\
\midrule
Overall & --- & 200 & 5.93 & 0.41 \\
\bottomrule
\end{tabular}
\end{table}

\begin{table*}[t]
\centering
\setlength{\abovecaptionskip}{5pt}
\setlength{\belowcaptionskip}{0pt}
\caption{Vulnerability severity and distribution across OWASP categories. Each vulnerability category is decomposed into Frontend, Backend, and Configuration components. 
\emph{Critical}, \emph{High}, \emph{Medium}, and \emph{Low} report the vulnerability number at each severity level within that component. \emph{Total} reports the total vulnerability number in the category and its share. 
\emph{Repo} reports the incidence rate of audited repositories containing at least one vulnerability in the category, with the corresponding rank shown as (\#). \emph{OWASP} reports OWASP's application-level incidence rate~\cite{owasp2025datafactors}, with the corresponding rank shown as (\#). The OWASP incidence values are provided for reference only and are not intended for direct comparison.}
\label{tab:rq2-owasp-component}
\footnotesize
\setlength{\tabcolsep}{6pt}
\renewcommand{\arraystretch}{0.9}
\definecolor{heatblue}{HTML}{2C6EBD}

\begin{tabular}{@{}lcrrrrccc@{\hspace{\tabcolsep}}}
\toprule
\textbf{Category}
& \textbf{Component}
& \textbf{Critical}
& \textbf{High}
& \textbf{Medium}
& \textbf{Low}
& \textbf{Total}
& \textbf{Repo}
& \textbf{OWASP} \\
\midrule

\multirow{3}{*}{Broken Access Control}
& Frontend
& \cellcolor{heatblue!8}9
& \cellcolor{heatblue!15}43
& \cellcolor{heatblue!30}21
& \cellcolor{heatblue!10}4
& \multirow{3}{*}{339 (28.58\%)}
& \cellcolor{heatblue!55}
& \cellcolor{heatblue!36} \\
& Backend
& \cellcolor{heatblue!55}61
& \cellcolor{heatblue!55}154
& \cellcolor{heatblue!55}39
& \cellcolor{heatblue!18}7
& & \cellcolor{heatblue!55}
& \cellcolor{heatblue!36} \\
& Configuration
& \cellcolor{heatblue!1}1
& 0
& 0
& 0
& & \multirow{-3}{*}{\cellcolor{heatblue!55}64.00\% (\#1)}
& \multirow{-3}{*}{\cellcolor{heatblue!36}3.74\% (\#4)} \\
\hline
\addlinespace[1pt]

\multirow{3}{*}{Injection}
& Frontend
& \cellcolor{heatblue!15}17
& \cellcolor{heatblue!19}52
& \cellcolor{heatblue!24}17
& \cellcolor{heatblue!22}9
& \multirow{3}{*}{199 (16.78\%)}
& \cellcolor{heatblue!46}
& \cellcolor{heatblue!30} \\
& Backend
& \cellcolor{heatblue!14}15
& \cellcolor{heatblue!22}61
& \cellcolor{heatblue!34}24
& \cellcolor{heatblue!5}2
& & \cellcolor{heatblue!46}
& \cellcolor{heatblue!30} \\
& Configuration
& 0
& \cellcolor{heatblue!1}2
& 0
& 0
& & \multirow{-3}{*}{\cellcolor{heatblue!46}54.00\% (\#2)}
& \multirow{-3}{*}{\cellcolor{heatblue!30}3.08\% (\#5)} \\
\hline
\addlinespace[1pt]

\multirow{3}{*}{Authentication Failures}
& Frontend
& \cellcolor{heatblue!17}19
& \cellcolor{heatblue!10}27
& \cellcolor{heatblue!35}25
& \cellcolor{heatblue!20}8
& \multirow{3}{*}{176 (14.84\%)}
& \cellcolor{heatblue!44}
& \cellcolor{heatblue!28} \\
& Backend
& \cellcolor{heatblue!23}26
& \cellcolor{heatblue!18}49
& \cellcolor{heatblue!10}7
& \cellcolor{heatblue!7}3
& & \cellcolor{heatblue!44}
& \cellcolor{heatblue!28} \\
& Configuration
& \cellcolor{heatblue!2}2
& \cellcolor{heatblue!1}2
& \cellcolor{heatblue!7}5
& \cellcolor{heatblue!7}3
& & \multirow{-3}{*}{\cellcolor{heatblue!44}51.50\% (\#3)}
& \multirow{-3}{*}{\cellcolor{heatblue!28}2.92\% (\#8)} \\
\hline
\addlinespace[1pt]

\multirow{3}{*}{Cryptographic Failures}
& Frontend
& \cellcolor{heatblue!14}16
& \cellcolor{heatblue!5}13
& \cellcolor{heatblue!21}15
& \cellcolor{heatblue!5}2
& \multirow{3}{*}{112 (9.44\%)}
& \cellcolor{heatblue!31}
& \cellcolor{heatblue!37} \\
& Backend
& \cellcolor{heatblue!7}8
& \cellcolor{heatblue!6}18
& \cellcolor{heatblue!10}7
& 0
& & \cellcolor{heatblue!31}
& \cellcolor{heatblue!37} \\
& Configuration
& \cellcolor{heatblue!25}28
& \cellcolor{heatblue!2}5
& 0
& 0
& & \multirow{-3}{*}{\cellcolor{heatblue!31}36.50\% (\#4)}
& \multirow{-3}{*}{\cellcolor{heatblue!37}3.80\% (\#3)} \\
\hline
\addlinespace[1pt]

\multirow{3}{*}{Security Misconfiguration}
& Frontend
& \cellcolor{heatblue!3}3
& \cellcolor{heatblue!2}6
& \cellcolor{heatblue!13}9
& \cellcolor{heatblue!2}1
& \multirow{3}{*}{108 (9.11\%)}
& \cellcolor{heatblue!31}
& \cellcolor{heatblue!29} \\
& Backend
& \cellcolor{heatblue!2}2
& \cellcolor{heatblue!13}37
& \cellcolor{heatblue!38}27
& \cellcolor{heatblue!7}3
& & \cellcolor{heatblue!31}
& \cellcolor{heatblue!29} \\
& Configuration
& \cellcolor{heatblue!6}7
& \cellcolor{heatblue!2}6
& \cellcolor{heatblue!8}6
& \cellcolor{heatblue!2}1
& & \multirow{-3}{*}{\cellcolor{heatblue!31}36.50\% (\#4)}
& \multirow{-3}{*}{\cellcolor{heatblue!29}3.00\% (\#6)} \\
\hline
\addlinespace[1pt]

\multirow{3}{*}{Supply Chain Failures}
& Frontend
& \cellcolor{heatblue!2}2
& \cellcolor{heatblue!1}4
& 0
& 0
& \multirow{3}{*}{73 (6.16\%)}
& \cellcolor{heatblue!24}
& \cellcolor{heatblue!55} \\
& Backend
& \cellcolor{heatblue!1}1
& 0
& \cellcolor{heatblue!1}1
& \cellcolor{heatblue!5}2
& & \cellcolor{heatblue!24}
& \cellcolor{heatblue!55} \\
& Configuration
& \cellcolor{heatblue!2}2
& \cellcolor{heatblue!13}36
& \cellcolor{heatblue!23}16
& \cellcolor{heatblue!22}9
& & \multirow{-3}{*}{\cellcolor{heatblue!24}28.00\% (\#7)}
& \multirow{-3}{*}{\cellcolor{heatblue!55}5.72\% (\#1)} \\
\hline
\addlinespace[1pt]

\multirow{3}{*}{Security Logging Failures}
& Frontend
& 0
& \cellcolor{heatblue!2}7
& \cellcolor{heatblue!17}12
& \cellcolor{heatblue!55}22
& \multirow{3}{*}{64 (5.40\%)}
& \cellcolor{heatblue!24}
& \cellcolor{heatblue!38} \\
& Backend
& 0
& \cellcolor{heatblue!1}2
& \cellcolor{heatblue!13}9
& \cellcolor{heatblue!25}10
& & \cellcolor{heatblue!24}
& \cellcolor{heatblue!38} \\
& Configuration
& 0
& \cellcolor{heatblue!1}1
& 0
& \cellcolor{heatblue!2}1
& & \multirow{-3}{*}{\cellcolor{heatblue!24}28.50\% (\#6)}
& \multirow{-3}{*}{\cellcolor{heatblue!38}3.91\% (\#2)} \\
\hline
\addlinespace[1pt]

\multirow{3}{*}{Insecure Design}
& Frontend
& 0
& \cellcolor{heatblue!3}8
& \cellcolor{heatblue!27}19
& \cellcolor{heatblue!20}8
& \multirow{3}{*}{62 (5.23\%)}
& \cellcolor{heatblue!21}
& \cellcolor{heatblue!18} \\
& Backend
& 0
& \cellcolor{heatblue!6}16
& \cellcolor{heatblue!16}11
& 0
& & \cellcolor{heatblue!21}
& \cellcolor{heatblue!18} \\
& Configuration
& 0
& 0
& 0
& 0
& & \multirow{-3}{*}{\cellcolor{heatblue!21}25.00\% (\#8)}
& \multirow{-3}{*}{\cellcolor{heatblue!18}1.86\% (\#10)} \\
\hline
\addlinespace[1pt]

\multirow{3}{*}{Exception Mishandling}
& Frontend
& 0
& \cellcolor{heatblue!1}2
& \cellcolor{heatblue!16}11
& \cellcolor{heatblue!22}9
& \multirow{3}{*}{41 (3.46\%)}
& \cellcolor{heatblue!14}
& \cellcolor{heatblue!28} \\
& Backend
& 0
& \cellcolor{heatblue!2}6
& \cellcolor{heatblue!13}9
& \cellcolor{heatblue!10}4
& & \cellcolor{heatblue!14}
& \cellcolor{heatblue!28} \\
& Configuration
& 0
& 0
& 0
& 0
& & \multirow{-3}{*}{\cellcolor{heatblue!14}16.00\% (\#9)}
& \multirow{-3}{*}{\cellcolor{heatblue!28}2.95\% (\#7)} \\
\hline
\addlinespace[1pt]

\multirow{3}{*}{Data Integrity Failures}
& Frontend
& 0
& \cellcolor{heatblue!1}1
& \cellcolor{heatblue!4}3
& \cellcolor{heatblue!5}2
& \multirow{3}{*}{12 (1.01\%)}
& \cellcolor{heatblue!5}
& \cellcolor{heatblue!26} \\
& Backend
& 0
& \cellcolor{heatblue!1}3
& \cellcolor{heatblue!1}1
& 0
& & \cellcolor{heatblue!5}
& \cellcolor{heatblue!26} \\
& Configuration
& 0
& 0
& \cellcolor{heatblue!1}1
& \cellcolor{heatblue!2}1
& & \multirow{-3}{*}{\cellcolor{heatblue!5}6.00\% (\#10)}
& \multirow{-3}{*}{\cellcolor{heatblue!26}2.75\% (\#9)} \\

\bottomrule
\end{tabular}
\end{table*}

\Cref{tab:size-groups} further shows that repository scale affects vulnerability prevalence in two different ways. In absolute terms, larger applications contain more vulnerabilities on average. The mean number of vulnerabilities per repository rises from 2.13 in Small applications to 4.31 in Medium, 6.28 in Large, and 12.38 in xLarge. In contrast, vulnerability density declines steadily as repository scale increases. Small applications have 0.85 vulnerabilities per 1{,}000 LoC, compared with 0.67 for Medium, 0.53 for Large, and 0.22 for xLarge applications. Thus, while larger applications accumulate more vulnerabilities overall, smaller applications contain vulnerabilities more densely relative to their code scales. This result suggests that small vibe-coded applications should not be assumed to be more secure simply because they contain less code.

\paragraph{Vulnerability Severity and Distribution.}
From \Cref{tab:rq2-owasp-component}, vibe-coded applications exhibit a high prevalence of severe vulnerabilities. Among the 1,186 identified vulnerabilities, 65.77\% are classified as Critical or High severity, including 18.47\% Critical and 47.30\% High. In contrast, only 9.36\% of vulnerabilities are classified as Low severity.

To further characterize these vulnerabilities, Table~\ref{tab:rq2-owasp-component} provides their distribution across OWASP categories.
Broken Access Control is the largest category, with 339 vulnerabilities (28.58\%) and affecting 64.00\% of vibe-coded applications. Injection and Authentication Failures are the next largest categories, with 199 (16.78\%) and 176 (14.84\%) vulnerabilities and affecting 54.00\% and 51.50\% of applications, respectively. Together, these three categories account for 60.20\% of all vulnerabilities, which aligns with failures at security-critical trust boundaries. Broken Access Control stems from missing or incomplete authorization checks; Injection captures cases where untrusted input reaches SQL, HTML, or other sensitive sinks without sufficient validation or sanitization; and Authentication Failures reflect missing or ineffective identity and session checks. These findings suggest that the main failures directly affect who can access protected resources, how users are authenticated, and how untrusted input flows through the application.

\begin{figure*}[t]
\centering
\setlength{\abovecaptionskip}{5pt}
\setlength{\belowcaptionskip}{-5pt}
\includegraphics[width=\textwidth]{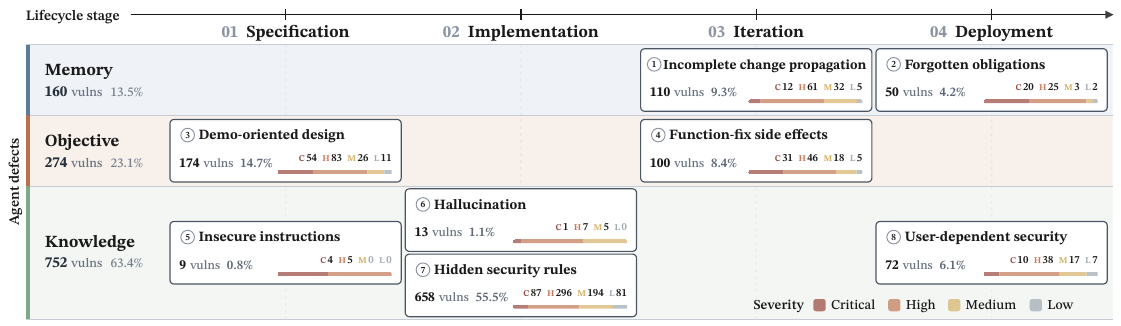}
\caption{Eight failure modes (\cnum{1}--\cnum{8}) for vibe-coded vulnerabilities, grouped by three defects (i.e., memory, objective, and knowledge) and mapped to the vibe-coding lifecycle (from specification and implementation, to iteration and deployment).}
\label{fig:rq3-lifecycle}
\end{figure*}

\Cref{tab:rq2-owasp-component} further shows that these vulnerabilities are unevenly distributed across implementation components. Overall, backend code accounts for 625 of the 1{,}186 vulnerabilities (52.70\%), compared with 426 (35.92\%) in frontend code and 135 (11.38\%) in configurations. This indicates that vulnerabilities are concentrated in server-side logic rather than being distributed evenly across the application stack. The component mix also differs across OWASP categories. Broken Access Control is backend-centric: 261 of its 339 vulnerabilities (76.99\%) appear in backend code, compared with 77 (22.71\%) in frontend code and only 1 (0.30\%) in configurations. By contrast, Injection is almost evenly split between backend and frontend code, with 102 of 199 vulnerabilities (51.26\%) in backend code and 95 (47.74\%) in frontend code. Similarly, for Authentication Failures, 79 of 176 vulnerabilities (44.89\%) occur in frontend code, and 85 (48.30\%) occur in backend code. Taken together, these results suggest that while backend code is the main location of vulnerabilities overall, different OWASP categories concentrate in different components.

We additionally report the OWASP Top 10 (2025) incidence statistics (OWASP) alongside our category distribution, as a contextual reference. Because the two datasets differ substantially in scope and detection methodology, this comparison can only be interpreted as a qualitative signal rather than a precise measurement. With this caveat in mind, our result suggests that Injection and Authentication Failures are relatively more prominent in our sample than in the OWASP reference distribution. This qualitative pattern is consistent with the hypothesis that vibe-coding workflows may underweight authorization checks and input validation at trust boundaries.

\begin{findingbox}[RQ.2 Findings]
Over 90\% of deployed vibe-coded web applications are insecure with at least one vulnerability, and 65.77\% of their vulnerabilities are classified as Critical or High severity. Vulnerabilities are concentrated in Broken Access Control, Injection, and Authentication Failures, which together account for 60.2\% of all vulnerabilities. Backend code constitutes the largest share of vulnerable components. 
\end{findingbox}

\section{RQ.3: What Factors Contribute to These Vulnerabilities?}
\label{sec:rq3}

To understand how vulnerabilities are introduced in vibe-coded applications, we qualitatively analyze each vulnerability. For each case, two authors independently examine the vulnerable code, the commit history that produced it, and any surrounding inline comments. Each author then assigns an open-ended label that describes how the vulnerability was introduced, instead of forcing cases into a predefined taxonomy. The initial labels showed substantial agreement between the two authors (Cohen's $\kappa = 0.82$). The authors subsequently consolidated their labels through discussion, merging overlapping tags and refining boundary cases until no new failure-mode categories emerged from additional cases. This process identified eight recurring \emph{failure modes} and grouped them into three broader \emph{defect classes}: \reasonM{memory defect}, \reasonO{objective defect}, and \reasonK{knowledge defect}, as illustrated in \Cref{fig:rq3-lifecycle}.

Furthermore, we map the failure modes into the vibe-coding lifecycle, which comprises four development stages. In the \emph{specification} stage, the user provides instructions, and the agent plans the solution architecture. In the \emph{implementation} stage, the agent generates code for the intended design. In the \emph{iteration} stage, the user and agent repeatedly test, debug, and revise the codebase. In the \emph{deployment} stage, the application is released to production and exposed to real users. These stages are not strictly linear and may recur as the user and agent repeatedly refine the application. We also report the number and severity of vulnerabilities in each failure mode. We next detail each failure mode using concrete real-world examples from our dataset.

\subsection{Memory Defects}\label{sec:memory}

\reasonM{Memory defects} arise when security obligations introduced earlier in development are not reliably retained, propagated, or revisited as the project evolves. Although coding agents can reconstruct context from prompts, files, and recent interactions, they do not maintain a complete and reliable memory of all project-wide security requirements over time. In the vibe-coding lifecycle, these defects are primarily in later-stage development, especially during \emph{iteration}, when security-relevant changes must be propagated consistently across the codebase, and \emph{deployment}, when deferred obligations must still be completed before release. Together, these defects account for 160 (13.5\%) vulnerabilities.

\paragraph{Iteration Stage: \cnum{1}\reasonM{incomplete change propagation}.} 
This failure mode occurs when the agent applies a security requirement in one part of the codebase but fails to propagate it to other relevant locations. Sometimes, this manifests as a ``forward propagation failure'', where a newly introduced security property is applied only to new code and never back-fitted to pre-existing paths. In other cases, it manifests as a ``backward conformance failure'', where a new code path is introduced that bypasses a security property already enforced throughout the rest of the codebase. The following code snippet illustrates the forward variant. A \texttt{validateApiKey} middleware is added for a newly introduced router, but the same protection is not applied to 11 pre-existing handlers. As a result, those handlers remain reachable without the intended authorization check. We identify 110 vulnerabilities (9.3\%) caused by this failure mode. Their severity is also substantial. 66.36\% are rated Critical or High. These vulnerabilities are especially common in Broken Access Control, Injection and Authentication Failures, where the same protection often needs to be applied across multiple endpoints, handlers, or sinks.

{
\setlength{\intextsep}{3pt}
\begin{figure}[H]
\centering
\includegraphics[width=0.92\columnwidth]{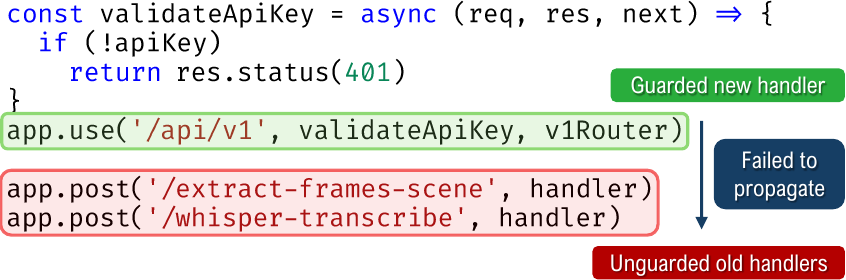}
\label{lst:propagation}
\end{figure}
}

\paragraph{Deployment Stage: \cnum{2}\reasonM{forgotten obligations}.} 
This failure mode occurs when a security-relevant task is explicitly marked as temporary, incomplete, or deferred during development, but is never revisited before deployment. The code may contain a TODO, FIXME, or other indication that the logic is not yet complete. As development progresses, the obligation may fall out of the active context, and no external mechanism ensures that it is resolved before release. This results in temporary code shipped in the same unfinished state. In the following example, the log-in route leaves password verification as a TODO because the password field has not yet been added to the database. This task is never completed, and the deployed application accepts incomplete authentication logic. Forgotten obligations account for 50 vulnerabilities (4.2\%), where 90.0\% are rated Critical or High. They concentrate on Broken Access Control and Authentication Failures, where temporary logic takes the form of authentication or authorization bypass.

{
\setlength{\intextsep}{3pt}
\begin{figure}[H]
\centering
\includegraphics[width=0.9\columnwidth]{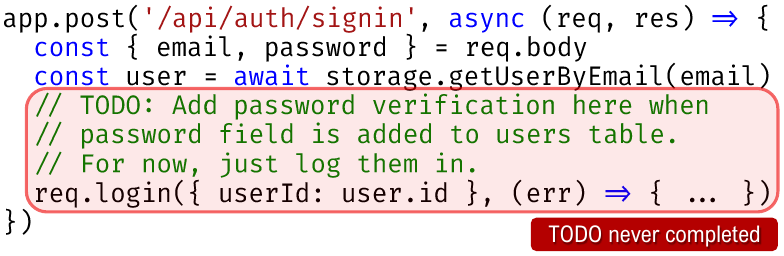}
\label{lst:forgotten}
\end{figure}
}

\subsection{Objective Defects}~\label{sec:objective}

\reasonO{Objective defects} occur when immediate functional goals are prioritized over long-term correctness or security during development. In these cases, security is systematically subordinated to short-term objectives such as making the application demo-ready, unblocking a visible error, or restoring functionality as quickly as possible. Objective defects are most prevalent in the \emph{specification} and \emph{iteration} stages. In total, they account for 274 (23.1\%) vulnerabilities.

\paragraph{Specification Stage: \cnum{3}\reasonO{demo-oriented design}.}
This failure mode occurs when the application is designed around immediate usability, convenience, or demonstrability rather than secure-by-default practice. In these cases, insecure mechanisms are not merely left to be fixed later. Instead, they are incorporated into the intended design. Examples include dummy encryption functions, static demo keys, or placeholder authorization logic that remains acceptable as long as the application appears to work. The following example illustrates this pattern. The application stores secret OAuth tokens in front-end \texttt{localStorage} after applying only base64 encoding, with the choice described as suitable for demonstration. This exposes sensitive credentials to client-side access. Demo-oriented design accounts for 174 vulnerabilities (14.7\%), mainly in Broken Access Control, Authentication Failures, and Cryptographic Failures. These vulnerabilities often lead to authentication bypasses or secret exposure, and 78.74\% are classified as Critical or High.

{
\setlength{\intextsep}{3pt}
\begin{figure}[H]
\centering
\includegraphics[width=0.92\columnwidth]{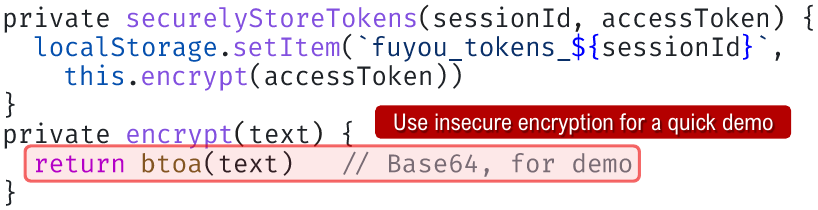}
\label{lst:demo}
\end{figure}
}

\paragraph{Iteration stage: \cnum{4}\reasonO{function-fix side effects}.}
This failure mode arises during debugging or feature iteration, when a visible error is resolved by weakening or bypassing a security control. For example, an authentication error, failed CORS preflight, or blocked operation may be addressed by relaxing the security boundary check. These insecure fixes appear to make the application work, so they often remain in the codebase. The following is an example: a bypass-login route with hard-coded credentials is introduced to work around a Supabase authentication error, creating a critical authentication vulnerability. Function-fix side effects lead to 100 vulnerabilities (8.4\%), mainly in Broken Access Control. Like demo-oriented design, this frequently produces high-impact failures. 77.0\% of its vulnerabilities are Critical or High.

{
\setlength{\intextsep}{3pt}
\begin{figure}[H]
\centering
\includegraphics[width=0.9\columnwidth]{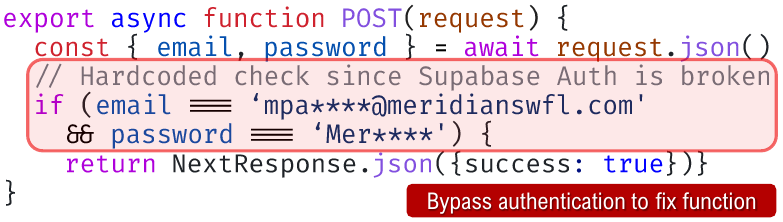}
\label{lst:funcfix}
\end{figure}
}

\subsection{Knowledge Defects}~\label{sec:knowledge}

\reasonK{Knowledge defects} refer to the fact that both agents and users may fail to recognize the security requirements of the application during its development. In the vibe-coding workflow, many security constraints are not explicitly stated, and agents may not derive them from the surrounding application context. On the user side, developers may overlook necessary security steps. These knowledge gaps on both sides result in insufficient, incorrect, or fabricated security practices.
This is the largest defect class, accounting for 752 vulnerabilities (63.4\%). Knowledge defects span the \emph{specification}, \emph{implementation}, and \emph{deployment} stages.

\paragraph{Specification Stage: \cnum{5}\reasonK{insecure instructions}.} 
This captures cases where the specification itself contains an insecure requirement, and the implementation follows that requirement faithfully. Unlike hidden security rules, where a necessary security property is omitted, insecure instructions explicitly direct development toward an unsafe design. The following shows an example in which \texttt{CLAUDE.md} instructs to embed an OpenAI API key in the code and ``protect'' it using a Caesar cipher. The agent implements this faithfully, resulting in the exposure of the secret. This mode accounts for 9 vulnerabilities (0.8\%), all of which are classified as Critical or High. This number is low because it is only observed when the insecure instruction is preserved in the repository or referenced by surrounding comments.

{
\setlength{\intextsep}{3pt}
\begin{figure}[H]
\centering
\includegraphics[width=0.8\columnwidth]{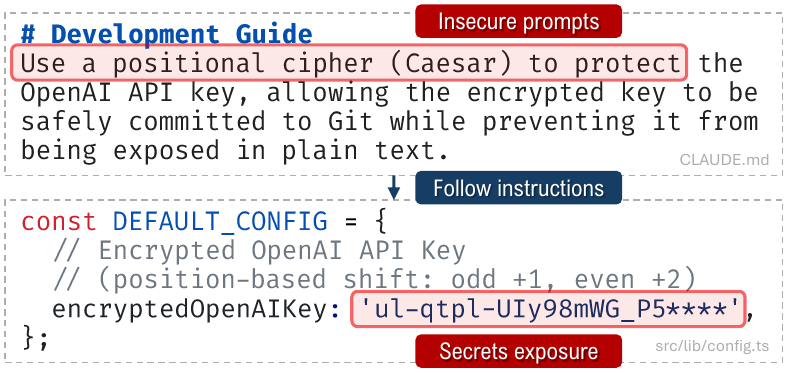}
\label{lst:insecure}
\end{figure}
}

\paragraph{Implementation Stage: \cnum{6}\reasonK{hallucination}.} 
This failure mode captures cases where the generated code relies on fabricated technical assumptions in security-sensitive logic. The assumption may concern a library API, a database schema, or the security property of a primitive. Many hallucination-induced mistakes break functionality and are therefore corrected during development. The cases that remain in our dataset are those that do not immediately cause an obvious error, allowing the application to keep running while the intended security mechanism is ineffective. The following illustrates this failure. The code invokes \texttt{crypto.createCipherGCM} as the encryption primitive, but this API does not exist in Node's \texttt{crypto} module. The encryption call therefore fails, and the surrounding error handling allows the secrets store to continue operating without effective encryption. Hallucination accounts for 13 vulnerabilities (1.1\%) in our dataset. Though this is a small category, it is analytically important as the code may implement a security mechanism while failing to provide the intended protection.

{
\setlength{\intextsep}{3pt}
\begin{figure}[H]
\centering
\includegraphics[width=0.8\columnwidth]{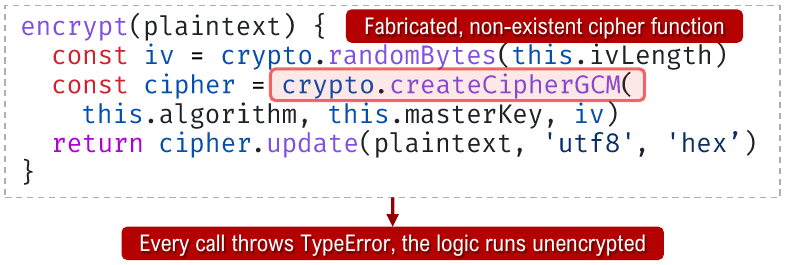}
\label{lst:hallucination}
\end{figure}
}

\paragraph{Implementation Stage: \cnum{7}\reasonK{hidden security rules}.}
This captures cases where generated code violates a security requirement that is necessary for safe implementation but is not explicit in the user request. Such requirements include hashing credentials before storage, validating input before sensitive sinks, and enforcing security checks on the server side. Since these requirements often do not affect visible functionality, their absence may go unnoticed. The following shows a representative example. The vulnerable code serializes an external response and writes it into the page using \texttt{document.write}, but omits the required escaping step, resulting in reflected XSS. Hidden security rules account for 658 vulnerabilities (55.5\%), making it the most common reason in our analysis. These vulnerabilities appear most frequently in Broken Access Control and Injection, where a single missing security requirement can affect many routes, handlers, or sinks. Although the violated rules vary in immediate impact, 58.21\% of these vulnerabilities are classified as Critical or High.

{
\setlength{\intextsep}{3pt}
\begin{figure}[H]
\centering
\includegraphics[width=0.7\columnwidth]{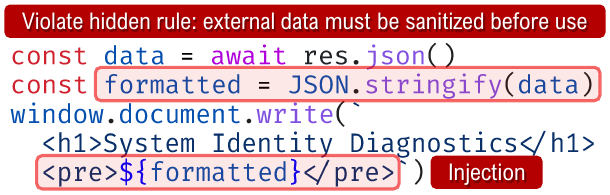}
\label{lst:hiddenrules}
\end{figure}
}

\paragraph{Deployment stage: \cnum{8}\reasonK{user-dependent security}.} 
This failure occurs when the vibe-coded application depends on a security-sensitive deployment or maintenance step that is left to the user. Examples include excluding generated secret files from Git, configuring necessary environment variables, and updating vulnerable dependencies. These may be routine for experienced developers, but vibe coding does not make the obligation explicit or enforced, creating a responsibility gap between what the agent assumes will happen and what the user is actually aware of. In the following example, the application exports database contents into files for backup and implicitly expects the user to exclude these files from Git. However, the user directly commits them to a public GitHub repository. This mode leads to 72 vulnerabilities (6.1\%), most commonly Supply Chain Failures due to exposed secrets, committed environment files, or unrotated default credentials. 66.67\% of vulnerabilities are classified as Critical or High.

{
\setlength{\intextsep}{3pt}
\begin{figure}[H]
\centering
\includegraphics[width=0.9\columnwidth]{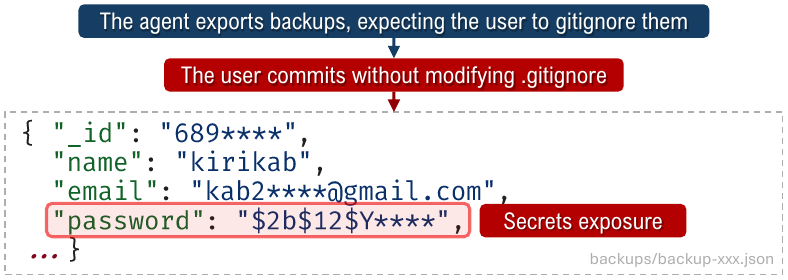}
\label{lst:userdep}
\end{figure}
}

\begin{findingbox}[RQ.3 Findings]
Insecurities in vibe-coded applications largely stem from systematic limitations in how agents retain, prioritize, and reason about security requirements across the development lifecycle. We identify eight recurring failure modes across three defect classes: memory, objective, and knowledge, with knowledge defects (63.4\%) the most prevalent.
\end{findingbox}

\section{RQ.4: What Conditions Can Mitigate Vulnerabilities?}
\label{sec:rq4}

To understand when vibe coding is likely to mitigate vulnerabilities, we study three potential conditions that may influence its behavior: (1) agent harness, (2) underlying language model, and (3) user prompt. To isolate the effect of each condition, we conduct a controlled replay experiment that re-executes the original vibe-coding scenarios in which vulnerabilities were introduced and measures whether the agent reintroduces them under a different configuration.

\begin{table}[t]
\centering
\caption{Eight experimental configurations in the OFAT design (each changes one factor relative to the baseline).}
\label{tab:rq4-configs}
\footnotesize
\setlength{\tabcolsep}{3pt}
\renewcommand{\arraystretch}{0.9}
\begin{tabular}{@{}l l >{\raggedright\arraybackslash}p{4.9cm}@{}}
\toprule
\textbf{Condition} & \textbf{Configuration} & \textbf{Setting} \\
\midrule
\textit{Baseline} & Baseline & default harness, GPT-5.6-Terra, short prompt \\
\midrule
Agent harness & Hardened & default plus a security-hardening skill~\cite{addyosmaniSecuritySkill} \\
\midrule
\multirow{2}{*}[-2pt]{Model}
          & Luna & weaker model \\
\cmidrule(l){2-3}
          & Sol  & stronger model \\
\midrule
\multirow{4}{*}[-9pt]{User prompt}
          & Professional & detailed prompt with technical instructions \\
\cmidrule(l){2-3}
          & Polished     & baseline plus ``write well-structured code'' \\
\cmidrule(l){2-3}
          & Production   & baseline plus ``ready for production'' \\
\cmidrule(l){2-3}
          & Self-check   & baseline plus ``review your changes'' \\
\bottomrule
\end{tabular}
\end{table}

\paragraph{Experimental Setup.}

We first select the vulnerability targets to replay. For this purpose, we identify the vulnerabilities in \vibevulns that are feasible to replay: the repository state before vulnerability introduction is available, the original task can be reconstructed faithfully, and the security outcome can be explicitly judged. We then stratify this replayable corpus by failure mode and randomly select 70 vulnerabilities. The resulting sample spans seven failure modes across all three defect classes, with multiple targets in every stratum. This includes 17 targets with \reasonM{memory defects}, 23 with \reasonO{objective defects}, and 30 with \reasonK{knowledge defects}. We exclude only \reasonK{hallucination}, as it is difficult to reproduce reliably.

For each target, we reconstruct the pre-state in which the vulnerability has not been introduced yet, while the surrounding feature context remains intact. To achieve this, we either roll back to the parent commit or surgically remove only the code change that introduces the vulnerability when the original commit contained significant feature changes that affected the whole context. After that, we manually issue a task prompt that requests the same functionality as the original vulnerability-introducing scenario.
We then run the agent to complete the task and collect both the generated artifacts and traces of the agentic process. 
Using these artifacts, we manually evaluate whether the target vulnerability has been reintroduced. We flag it as triggered if the same or similar vulnerability reappears; otherwise not. 

To investigate how these conditions affect vulnerability introduction, we adopt a one-factor-at-a-time (OFAT) design in which each run differs from the baseline by exactly one factor, as shown in \Cref{tab:rq4-configs}. The baseline uses the default agent harness, the mid-capability \emph{GPT-5.6-Terra} model, and a short non-expert prompt. We then vary each of the three factors. For the agent harness, we compare the default with that augmented with a security-hardening skill~\cite{addyosmaniSecuritySkill}. For the underlying model, we compare GPT-5.6-Terra with the weaker \emph{GPT-5.6-Luna} and the stronger \emph{GPT-5.6-Sol}. For user prompts, we apply four strategies to the baseline prompt: (1) making it more professional and detailed, (2) requesting a polished implementation, (3) specifying that the code should be production-ready, and (4) asking the agent to perform a self-check. All prompts request identical functionality and do not reveal the target security criterion. This yields eight configurations per target. We repeat each configuration three times to mitigate the stochasticity of agent behaviors. As a result, we have $70 \times 8 \times 3 = 1{,}680$ controlled runs.

\paragraph{Analysis Results.} The results are shown in \Cref{tab:rq4-reason-config}. Overall, the target vulnerability is reintroduced in 434 of 1{,}680 runs (25.8\%). For the baseline, it is reintroduced in 54 of 210 runs (25.7\%), showing that vulnerability reintroduction is not limited to isolated one-time outcomes.

\begingroup
\definecolor{rq4Decrease}{HTML}{72B89A}
\definecolor{rq4Increase}{HTML}{D98278}
\newcommand{\rqbase}[1]{#1}
\newcommand{\rqflat}[1]{#1}
\newcommand{\rqdown}[2]{\cellcolor{rq4Decrease!#1}#2}
\newcommand{\rqup}[2]{\cellcolor{rq4Increase!#1}#2}
\begin{table*}[t]
\centering
\caption{Vulnerability reintroduction rates by defect class, failure mode, and OFAT configuration in the controlled replay experiment. Each cell reports the number of triggered replay runs over evaluated runs, followed by the corresponding percentage. $n_t$ denotes the number of vulnerability targets, with each target contributing three repeats per configuration. The bottom row reports the change in the overall reintroduction rate for each configuration relative to the baseline ($\Delta$ vs. Base). Green indicates a decrease, red an increase, and white little or no change; darker shading denotes a larger absolute change.}
\label{tab:rq4-reason-config}
\footnotesize
\setlength{\tabcolsep}{0pt}
\renewcommand{\arraystretch}{1.2}
\resizebox{0.99\textwidth}{!}{\begin{tabular}{@{}llrrrrrrrrr@{}}
\toprule
\textbf{Defect} & \textbf{Failure Mode} & $\boldsymbol{n_t}$ & \textbf{Base} & \textbf{Hardened} & \textbf{Luna} & \textbf{Sol} & \textbf{Polished} & \textbf{Production} & \textbf{Professional} & \textbf{Self-check} \\
\midrule
\multirow{3}{*}{\reasonM{Memory}}
 & \reasonM{Forgotten obligations} & 7 & \rqbase{6/21 (28.6\%)} & \rqflat{6/21 (28.6\%)} & \rqup{40}{9/21 (42.9\%)} & \rqflat{6/21 (28.6\%)} & \rqflat{6/21 (28.6\%)} & \rqdown{40}{3/21 (14.3\%)} & \rqup{60}{11/21 (52.4\%)} & \rqflat{6/21 (28.6\%)} \\
 & \reasonM{Incomplete change prop.} & 10 & \rqbase{4/30 (13.3\%)} & \rqdown{35}{1/30 (3.3\%)} & \rqup{20}{6/30 (20.0\%)} & \rqup{20}{6/30 (20.0\%)} & \rqdown{20}{2/30 (6.7\%)} & \rqdown{35}{1/30 (3.3\%)} & \rqup{45}{9/30 (30.0\%)} & \rqdown{15}{3/30 (10.0\%)} \\
 & \textit{Defect aggregate} & \textbf{17} & \rqbase{\textbf{10/51 (19.6\%)}} & \rqdown{20}{\textbf{7/51 (13.7\%)}} & \rqup{30}{\textbf{15/51 (29.4\%)}} & \rqup{15}{\textbf{12/51 (23.5\%)}} & \rqdown{15}{\textbf{8/51 (15.7\%)}} & \rqdown{35}{\textbf{4/51 (7.8\%)}} & \rqup{50}{\textbf{20/51 (39.2\%)}} & \rqflat{\textbf{9/51 (17.6\%)}} \\
\midrule
\multirow{3}{*}{\reasonO{Objective}}
 & \reasonO{Demo-oriented design} & 14 & \rqbase{13/42 (31.0\%)} & \rqdown{50}{5/42 (11.9\%)} & \rqup{35}{18/42 (42.9\%)} & \rqdown{20}{10/42 (23.8\%)} & \rqdown{20}{10/42 (23.8\%)} & \rqdown{50}{5/42 (11.9\%)} & \rqup{60}{23/42 (54.8\%)} & \rqdown{15}{11/42 (26.2\%)} \\
 & \reasonO{Function-fix side effects} & 9 & \rqbase{13/27 (48.1\%)} & \rqdown{55}{7/27 (25.9\%)} & \rqdown{20}{11/27 (40.7\%)} & \rqdown{35}{10/27 (37.0\%)} & \rqdown{40}{9/27 (33.3\%)} & \rqdown{55}{7/27 (25.9\%)} & \rqup{20}{15/27 (55.6\%)} & \rqdown{20}{11/27 (40.7\%)} \\
 & \textit{Defect aggregate} & \textbf{23} & \rqbase{\textbf{26/69 (37.7\%)}} & \rqdown{55}{\textbf{12/69 (17.4\%)}} & \rqup{15}{\textbf{29/69 (42.0\%)}} & \rqdown{25}{\textbf{20/69 (29.0\%)}} & \rqdown{30}{\textbf{19/69 (27.5\%)}} & \rqdown{55}{\textbf{12/69 (17.4\%)}} & \rqup{50}{\textbf{38/69 (55.1\%)}} & \rqdown{20}{\textbf{22/69 (31.9\%)}} \\
\midrule
\multirow{5}{*}{\reasonK{Knowledge}}
 & \reasonK{Hidden security rules} & 22 & \rqbase{16/66 (24.2\%)} & \rqdown{45}{6/66 (9.1\%)} & \rqup{45}{27/66 (40.9\%)} & \rqflat{16/66 (24.2\%)} & \rqflat{16/66 (24.2\%)} & \rqdown{50}{4/66 (6.1\%)} & \rqup{35}{24/66 (36.4\%)} & \rqflat{16/66 (24.2\%)} \\
 & \reasonK{Insecure instructions} & 4 & \rqbase{1/12 (8.3\%)} & \rqdown{25}{0/12 (0.0\%)} & \rqup{70}{8/12 (66.7\%)} & \rqup{60}{4/12 (33.3\%)} & \rqup{65}{5/12 (41.7\%)} & \rqup{45}{3/12 (25.0\%)} & \rqup{70}{7/12 (58.3\%)} & \rqup{45}{3/12 (25.0\%)} \\
 & \reasonK{User-dependent security} & 4 & \rqbase{1/12 (8.3\%)} & \rqdown{25}{0/12 (0.0\%)} & \rqup{60}{4/12 (33.3\%)} & \rqup{45}{3/12 (25.0\%)} & \rqdown{25}{0/12 (0.0\%)} & \rqdown{25}{0/12 (0.0\%)} & \rqup{70}{7/12 (58.3\%)} & \rqdown{25}{0/12 (0.0\%)} \\
 & \textit{Defect aggregate} & \textbf{30} & \rqbase{\textbf{18/90 (20.0\%)}} & \rqdown{40}{\textbf{6/90 (6.7\%)}} & \rqup{60}{\textbf{39/90 (43.3\%)}} & \rqup{20}{\textbf{23/90 (25.6\%)}} & \rqup{15}{\textbf{21/90 (23.3\%)}} & \rqdown{35}{\textbf{7/90 (7.8\%)}} & \rqup{60}{\textbf{38/90 (42.2\%)}} & \rqflat{\textbf{19/90 (21.1\%)}} \\
\midrule
\multicolumn{2}{l}{\textbf{Overall}} & \textbf{70} & \rqbase{\textbf{54/210 (25.7\%)}} & \rqdown{45}{\textbf{25/210 (11.9\%)}} & \rqup{45}{\textbf{83/210 (39.5\%)}} & \rqflat{\textbf{55/210 (26.2\%)}} & \rqdown{15}{\textbf{48/210 (22.9\%)}} & \rqdown{50}{\textbf{23/210 (11.0\%)}} & \rqup{60}{\textbf{96/210 (45.7\%)}} & \rqflat{\textbf{50/210 (23.8\%)}} \\
\multicolumn{3}{l}{\textbf{$\Delta$ vs. Base}} & \rqbase{\textbf{0.0 pp}} & \rqdown{45}{\textbf{$-$13.8 pp}} & \rqup{45}{\textbf{+13.8 pp}} & \rqflat{\textbf{+0.5 pp}} & \rqdown{15}{\textbf{$-$2.9 pp}} & \rqdown{50}{\textbf{$-$14.8 pp}} & \rqup{60}{\textbf{+20.0 pp}} & \rqflat{\textbf{$-$1.9 pp}} \\
\bottomrule
\end{tabular}}
\end{table*}
\endgroup

\textit{Effect of Each Factor:} Overall, most investigated conditions mitigate vulnerabilities to varying degrees. Compared with the baseline, the production-ready prompt achieves the largest reduction (14.8 pp), followed closely by the hardened harness (13.8 pp). This indicates that security guidance is effective at both the workflow level, through reusable security checks, and the task level, by making robustness an explicit completion criterion. Other prompting strategies are less impactful, reducing vulnerability reintroduction by only 2.9 pp (polished prompt) and 1.9 pp (self-check prompt). 
The professional prompt even raises it by 20.0 pp. Based on our trace observations, the detailed technical instructions in the professional prompt cause the agent to follow the requested specification more literally, leaving less room for its own security judgment.
Stronger models also improve security, reducing vulnerability reintroduction by 13.8 pp from Luna to Terra. However, gains diminish and even reverse by 0.5 pp from Terra to Sol, suggesting that scaling model capability alone is not a reliable path to improved security.

\textit{Effect by Defect Class and Failure Mode:} At the defect-class level, the interventions show distinct strengths. Compared with Terra, Sol reduces \reasonO{objective defects} by 8.7 pp but increases \reasonM{memory defects} and \reasonK{knowledge defects} by 3.9 and 5.6 pp. Likewise, polished and self-check prompts reduce \reasonO{objective defects} by 10.1 and 5.8 pp but do not improve \reasonK{knowledge defects}, which improve only under the hardened harness and production-ready prompt. Broad quality cues therefore appear more effective against insecure prioritisation than against missing security requirements. The failure-mode results clarify this pattern. The hardened harness leaves \reasonM{forgotten obligations} unchanged at 28.6\% but reduces \reasonM{incomplete change propagation} from 13.3\% to 3.3\%, suggesting that reusable checks catch inconsistent protections more readily than deferred tasks. Meanwhile, the production-ready prompt reduces \reasonK{hidden security rules} from 24.2\% to 6.1\% but raises \reasonK{insecure instructions} from 8.3\% to 25.0\%. It may thus surface omitted requirements without overriding an explicitly insecure request. Overall, effective mitigation depends on the source of the defect.

\textit{Persistence and Stochasticity:} Vulnerabilities remain persistent under the evaluated conditions. 35 vulnerabilities are triggered under at least three of the eight configurations, and no configuration eliminates all target vulnerabilities. Agent behavior is also stochastic. Of the 560 vulnerability--configuration cells, 123 (22.0\%) are non-unanimous across their three identical repeats, indicating that the same task, repository state, and configuration can produce both secure and insecure outcomes. This suggests that vibe coding security is inherently variable.

\textit{Security Awareness--Action Gap:} We further examine whether agents possess the security knowledge needed to recognize the vulnerabilities they introduce and whether they act on that knowledge. We manually review the execution traces of triggered runs and identify cases in which the agent explicitly acknowledges the associated security risk. In 90 of 434 triggered runs (20.7\%), the agent recognizes the risk yet still produces and ships insecure code. Rather than implementing a secure solution, it often substitutes warnings, comments, or recommendations for corrective action. This awareness--action gap shows that recognizing a security requirement is not sufficient unless the agent is also required to enforce it before task completion.

\begin{findingbox}[RQ.4 Findings]
Nearly all conditions, including hardened agent harness, enhanced underlying models, and carefully-crafted prompts, can help mitigate vibe-coding vulnerabilities to varying degrees. Among them, a production-ready prompt and a hardened agent harness are the most effective, and vulnerabilities with different defects respond differently to the conditions.
Furthermore, our analysis reveals a security awareness--action gap: agents often recognize security risks but fail to translate this awareness into secure implementation. 
\end{findingbox}
\section{Discussion}

\paragraph{Insecurity Sources.} Our findings suggest that insecurity in vibe coding arises from three interacting sources. 

\emph{Agent-level limitations:} Many of the observed defects can be traced to the behavior of the agent itself. First, agents do not reliably retain or propagate security-relevant context as the project evolves, which leaves earlier security properties unenforced in newly added or modified code. Second, agents tend to prioritize completing the immediate task over verifying the security implications of a proposed fix. Third, agents may lack, or fail to surface, security requirements that are not explicit in the user's request, and may occasionally rely on fabricated technical assumptions in security-sensitive logic. 

\emph{Workflow-level weaknesses:} Conventional software development workflows may also forget plans and write temporary code, but they mitigate these by requirement checks, code review, and release gates before deployment. In contrast, vibe-coded applications are typically developed in short, high-intensity bursts, leaving little room for such checks, and non-expert users also cannot provide safeguards on their own. As a result, vulnerabilities that would ordinarily be caught at distinct stages can propagate through the pipeline together.

\emph{Responsibility misalignment:} Vibe coding is asymmetric: the agent performs most of the design and implementation work, while the user often has minimal technical participation. However, many AI agents still assign responsibility for design quality, security review, deployment, and maintenance to the user. This creates a mismatch between responsibility and capability. Users who lack security expertise are not well positioned to judge secure design choices, detect security flaws in generated code, or maintain secure deployments, leaving vulnerabilities to emerge and persist unchecked.

\paragraph{Takeaways for Different Stakeholders.} Our findings point to distinct, actionable takeaways for the key stakeholders involved in the vibe-coding pipeline.

\emph{For agent and harness developers:} Recurring implicit requirements (e.g., server-side authorization) should be compiled into a checkable rule set and enforced at every stage of the agent lifecycle, akin to policy-as-code integrated into the agent harness. Developers should also address the awareness--action gap directly. In our study, agents explicitly recognized the risks yet still produced insecure code; therefore, unresolved risks should trigger a structured blocking mechanism rather than being relegated to warnings or code comments. Finally, before deployment, agents should automatically verify critical safeguards, such as excluding secret files and rotating default credentials. This shifts responsibility away from end users, who may lack the expertise to assess such risks.

\emph{For foundation model developers:} While scaling model capability is important, it does not provide a uniformly reliable path to secure-by-default behavior. While stronger models reduce the incidence of vulnerabilities (from GPT-5.6-Luna to GPT-5.6-Terra), they reach a point of diminishing returns (from GPT-5.6-Terra to GPT-5.6-Sol). Therefore, improving security-by-default behavior may require alignment interventions specifically designed to preserve security properties under such pressure.

\emph{For end users:} End users can mitigate security risks by optimizing the development workflow. In the specification stage, they should provide agents with explicit security instructions. In our study, requesting a production-ready implementation and installing a security-hardening skill substantially reduced the vulnerability-triggering rate. Conversely, users should avoid supplying technical details they do not fully understand, as detailed professional prompts can sometimes lead agents to follow insecure specifications too literally. During development and iteration, users should take security warnings seriously rather than approving immediately. Before public deployment, they should also request a self-check covering deferred tasks, authentication and authorization paths, input handling, and secrets. Nevertheless, no evaluated condition eliminated all vulnerabilities, so generated applications should be treated as prototypes and not casually exposed to real users.

\paragraph{Limitations and Threats to Validity.}

We identify several limitations that may threaten the validity and generalizability of our findings. First, to collect vibe-coded applications, we rely on fingerprints associated with two widely-used coding agents (Claude Code and Lovable), although some collected repositories show evidence of development using multiple agents. Consequently, our findings may not generalize to the entire vibe-coding landscape. To mitigate this threat, we analyze the two agent-associated repository groups separately. The main findings replicate across both groups: vulnerability prevalence is similarly high, the OWASP category rankings are correlated ($\rho=0.73$), and the root-cause rankings show strong concordance ($\rho=0.86$). Although several compositional differences remain, these results suggest that our main conclusions are not confined to repositories associated with a single agent.
Second, our dataset consists exclusively of open-source applications hosted on GitHub, and our findings may therefore be limited to this setting. Third, our vulnerability-detection framework may produce false positives or false negatives. To mitigate false positives, we conduct human validation of the detected vulnerabilities. We also evaluate our approach against human analysis and a set of known vulnerabilities. The results indicate that our framework is unlikely to miss a substantial number of vulnerabilities. Finally, we use Claude Code as one of our security auditors to discover vulnerabilities in applications generated by Claude Code, which is due to its strong capabilities and widespread adoption among vibe-coding users. This overlap may introduce a potential bias in vulnerability detection. To mitigate this, we employ two independent auditors to cross-validate results. We further validate our framework against human security analysis and a known vulnerability set, providing additional evidence that the results are unlikely to be driven by this potential bias. 
\section{Conclusion}
In this paper, we study the prevalence, severity, and contributing factors of vulnerabilities in vibe-coded applications, and investigate what conditions can mitigate them. Our study of 1,186 vulnerabilities from real-world, publicly deployed vibe-coded applications shows that (1) vulnerabilities are widespread, with 91.0\% of applications containing at least one vulnerability; (2) these vulnerabilities are often severe, with 65.77\% rated as Critical or High, concentrated in broken access control, injection, and authentication failures; and (3) they are traced to eight recurring failure modes across three defect classes---memory, objective, and knowledge defects---reflecting how security obligations are forgotten, deprioritized in favor of immediate functionality, or never recognized in the first place. We further show that, while better agent harnesses and prompts can reduce these risks to some extent, they do not eliminate them, and agents often recognize security risks yet still ship insecure code. Our study suggests that securing vibe-coded applications requires treating vibe coding as a pipeline-level security problem, rather than relying on code-generation quality alone.

\bibliographystyle{plainurl}
\bibliography{reference}

@article{se3.0,
author = {Hassan, Ahmed E. and Oliva, Gustavo A. and Lin, Dayi and Chen, Boyuan and Jiang, Zhen Ming (Jack)},
title = {Towards AI-Native Software Engineering (SE 3.0): A Vision and a Challenge Roadmap},
year = {2026},
publisher = {Association for Computing Machinery},
address = {New York, NY, USA},
issn = {1049-331X},
url = {https://doi.org/10.1145/3807901},
doi = {10.1145/3807901},
note = {Just Accepted},
journal = {ACM Trans. Softw. Eng. Methodol.},
month = apr
}

@article{secodeplt,
  title={SECODEPLT: A unified benchmark for evaluating the security risks and capabilities of code genAI},
  author={Nie, Yuzhou and Wang, Zhun and Yang, Yu and Jiang, Ruizhe and Tang, Yuheng and Davies, Xander and Gal, Yarin and Li, Bo and Guo, Wenbo and Song, Dawn},
  journal={Advances in Neural Information Processing Systems},
  volume={38},
  year={2026},
  url={https://arxiv.org/abs/2410.11096}
}

@misc{openaiSecuritySkill,
  author       = {{OpenAI}},
  title        = {{Security Best Practices Skill}},
  url          = {https://github.com/openai/skills/tree/main/skills/.curated/security-best-practices},
  note         = {Accessed: 2026-08-20}
}

@misc{nvd2025cve48757,
  author       = {{NVD}},
  title        = {{CVE-2025-48757}},
  url          = {https://nvd.nist.gov/vuln/detail/CVE-2025-48757},
  note         = {Accessed: 2026-08-20}
}

@misc{antigravitySkills,
  author       = {{sickn33}},
  title        = {{Antigravity Awesome Skills: Security Audit}},
  url          = {https://github.com/sickn33/antigravity-awesome-skills/blob/main/skills/security-audit/SKILL.md},
  note         = {Accessed: 2026-08-20}
}

@misc{juiceShop,
  author       = {{OWASP}},
  title        = {{OWASP Juice Shop}},
  url          = {https://owasp.org/www-project-juice-shop/},
  note         = {Accessed: 2026-08-20}
}

@misc{addyosmaniSecuritySkill,
  author       = {{Addy Osmani}},
  title        = {{Agent Skills: Security and Hardening}},
  url          = {https://github.com/addyosmani/agent-skills/blob/main/skills/security-and-hardening/SKILL.md},
  note         = {Accessed: 2026-08-20}
}

@misc{copilot,
  author       = {{Microsoft}},
  title        = {{GitHub Copilot---Your AI pair programmer}},
  url          = {https://github.com/features/copilot},
  note         = {Accessed: 2026-08-20}
}

@misc{codex,
  author       = {{OpenAI}},
  title        = {{Codex CLI}},
  url          = {https://github.com/openai/codex},
  note         = {Accessed: 2026-08-20}
}

@misc{claudeCode,
  author       = {{Anthropic}},
  title        = {{Claude Code}},
  url          = {https://claude.com/product/claude-code},
  note         = {Accessed: 2026-08-20}
}

@misc{lovable,
  author       = {{Lovable}},
  title        = {{Lovable}},
  url          = {https://lovable.dev/},
  note         = {Accessed: 2026-08-20}
}

@misc{lovableStatistics,
  author       = {{Panto}},
  title        = {{Lovable AI Statistics 2026---Users, Revenue, Adoption \& Market Metrics}},
  url          = {https://www.getpanto.ai/blog/lovable-statistics},
  note         = {Accessed: 2026-08-20}
}

@misc{karpathy2025vibe,
  author       = {Andrej Karpathy},
  title        = {There's a new kind of coding {I} call ``vibe coding''},
  year         = {2025},
  month        = feb,
  howpublished = {Post on X (formerly Twitter)},
  url          = {https://x.com/karpathy/status/1886192184808149383},
  note         = {Accessed: 2026-08-20}
}

@inproceedings{chou2025building,
  author        = {Yi-Hung Chou and Boyuan Jiang and Yi Wen Chen and Mingyue Weng and Victoria Jackson and Thomas Zimmermann and James A. Jones},
  title         = {Building Software by Rolling the Dice: A Qualitative Study of Vibe Coding},
  booktitle     = {Proceedings of the ACM Joint European Software Engineering Conference and Symposium on the Foundations of Software Engineering (ESEC/FSE)},
  year          = {2026},
  note          = {Accepted; to appear},
  eprint        = {2512.22418},
  archivePrefix = {arXiv},
  url           = {https://arxiv.org/abs/2512.22418}
}

@inproceedings{chandra2024aise,
  title={Ai in software engineering at google: Progress and the path ahead (invited talk)},
  author={Chandra, Satish},
  booktitle={Proceedings of the 1st ACM International Conference on AI-Powered Software},
  pages={182--182},
  year={2024}
}

@misc{shani2023github,
  author       = {Inbal Shani and {GitHub Staff}},
  title        = {Survey Reveals {AI}'s Impact on the Developer Experience},
  year         = {2023},
  howpublished = {GitHub Blog (Research)},
  url          = {https://github.blog/news-insights/research/survey-reveals-ais-impact-on-the-developer-experience/},
  note         = {Accessed: 2026-08-20}
}

@misc{vibegraveyard2025,
  author = {{Vibe Graveyard}},
  title  = {Vibe Graveyard: Real-World Security Incidents in Vibe-Coded Applications},
  year   = {2025},
  url    = {https://www.vibegraveyard.ai/},
  note   = {Accessed: 2026-04-29}
}

@misc{databricks2025passing,
  author = {{Databricks}},
  title  = {Passing the Security Vibe Check: The Dangers of Vibe Coding},
  year   = {2025},
  url    = {https://www.databricks.com/blog/passing-security-vibe-check-dangers-vibe-coding},
  note   = {Accessed: 2026-04-29}
}

@misc{owasp2025top10,
  author = {{OWASP Foundation}},
  title  = {{OWASP} Top 10 for {LLM} Applications 2025},
  year   = {2025},
  url    = {https://genai.owasp.org/resource/owasp-top-10-for-llm-applications-2025/},
  note   = {Accessed: 2026-04-29}
}

@misc{owasp2025webtop10,
  author = {{OWASP Top 10 Security Risk}},
  title  = {{OWASP} Top 10:2025---The Ten Most Critical Web Application Security Risks},
  year   = {2025},
  url    = {https://owasp.org/Top10/2025/},
  note   = {Accessed: 2026-05-21}
}

@misc{owasp2025datafactors,
  author = {{OWASP Data Factors}},
  title  = {{OWASP} Top 10:2025---What are Application Security Risks? (Data Factors)},
  year   = {2025},
  url    = {https://owasp.org/Top10/2025/0x02_2025-What_are_Application_Security_Risks/\#Data\%20Factors},
  note   = {Accessed: 2026-06-12}
}

@misc{owaspRiskRatingMethodology,
  author = {{OWASP Foundation}},
  title  = {{OWASP} Risk Rating Methodology},
  url    = {https://owasp.org/www-community/OWASP_Risk_Rating_Methodology},
  note   = {Accessed: 2026-06-12}
}

@misc{shahid2025llmcsec,
  author        = {Muhammad Usman Shahid and Chuadhry Mujeeb Ahmed and Rajiv Ranjan},
  title         = {{LLM-CSEC}: Empirical Evaluation of Security in {C/C++} Code Generated by Large Language Models},
  year          = {2025},
  eprint        = {2511.18966},
  archivePrefix = {arXiv},
  url           = {https://arxiv.org/abs/2511.18966}
}

@misc{lian2025ase,
  author        = {Keke Lian and Bin Wang and Lei Zhang and Libo Chen and Junjie Wang and Ziming Zhao and Yujiu Yang and others},
  title         = {{A.S.E}: A Repository-Level Benchmark for Evaluating Security in {AI}-Generated Code},
  year          = {2025},
  eprint        = {2508.18106},
  archivePrefix = {arXiv},
  url           = {https://arxiv.org/abs/2508.18106}
}

@inproceedings{schreiber2025security,
  title={Security vulnerabilities in ai-generated code: A large-scale analysis of public github repositories},
  author={Schreiber, Maximilian and Tippe, Pascal},
  booktitle={International Conference on Information and Communications Security},
  pages={153--172},
  year={2025},
  organization={Springer}
}

@article{tihanyi2025secure,
  author  = {Norbert Tihanyi and Tamas Bisztray and Mohamed Amine Ferrag and Ridhi Jain and Lucas C. Cordeiro},
  title   = {How Secure is {AI}-Generated Code: A Large-Scale Comparison of Large Language Models},
  journal = {Empirical Software Engineering},
  volume  = {30},
  number  = {2},
  pages   = {47},
  year    = {2025},
  doi     = {10.1007/s10664-024-10590-1},
  url     = {https://doi.org/10.1007/s10664-024-10590-1}
}

@inproceedings{dora2025hidden,
  title={The hidden risks of llm-generated web application code: A security-centric evaluation of code generation capabilities in large language models},
  author={Dora, Swaroop and Lunkad, Deven and Aslam, Naziya and Venkatesan, S and Shukla, Sandeep Kumar},
  booktitle={International Conference on Information Systems Security},
  pages={27--37},
  year={2025},
  url={https://arxiv.org/abs/2504.20612}
}

@InProceedings{schaad2025still,
author="Schaad, Andreas
and G{\"o}tz, Stefan
and Binder, Dominik",
editor="Nemec Zlatolas, Lili
and Rannenberg, Kai
and Welzer, Tatjana
and Garcia-Alfaro, Joaquin",
title="You Still have to Study On the Security of LLM Generated Code",
booktitle="ICT Systems Security and Privacy Protection",
year="2025",
pages="111--124",
url={https://link.springer.com/chapter/10.1007/978-3-031-92886-4_8}
}

@article{fu2025security,
author = {Fu, Yujia and Liang, Peng and Tahir, Amjed and Li, Zengyang and Shahin, Mojtaba and Yu, Jiaxin and Chen, Jinfu},
title = {Security Weaknesses of Copilot-Generated Code in GitHub Projects: An Empirical Study},
year = {2025},
issue_date = {November 2025},
publisher = {Association for Computing Machinery},
address = {New York, NY, USA},
volume = {34},
number = {8},
issn = {1049-331X},
url = {https://doi.org/10.1145/3716848},
doi = {10.1145/3716848},
journal = {ACM Trans. Softw. Eng. Methodol.},
month = oct,
articleno = {218},
numpages = {34}
}

@inproceedings{perry2023users,
  author    = {Perry, Neil and Srivastava, Megha and Kumar, Deepak and Boneh, Dan},
  title     = {Do Users Write More Insecure Code with {AI} Assistants?},
  booktitle = {Proceedings of the 2023 ACM SIGSAC Conference on Computer and Communications Security},
  series    = {CCS '23},
  pages     = {2785--2799},
  year      = {2023},
  month     = nov,
  doi       = {10.1145/3576915.3623157},
  url       = {https://doi.org/10.1145/3576915.3623157},
  publisher = {ACM}
}

@inproceedings{pearce2022asleep,
  author    = {Hammond Pearce and Baleegh Ahmad and Benjamin Tan and Brendan Dolan-Gavitt and Ramesh Karri},
  title     = {Asleep at the Keyboard? {Assessing} the Security of {GitHub Copilot}'s Code Contributions},
  booktitle = {2022 IEEE Symposium on Security and Privacy (SP)},
  pages     = {754--768},
  year      = {2022},
  doi       = {10.1109/SP46214.2022.9833571},
  url       = {https://doi.org/10.1109/SP46214.2022.9833571}
}

@INPROCEEDINGS{khoury2023secure,
  author={Khoury, Raphaël and Avila, Anderson R. and Brunelle, Jacob and Camara, Baba Mamadou},
  booktitle={2023 IEEE International Conference on Systems, Man, and Cybernetics (SMC)}, 
  title={How Secure is Code Generated by ChatGPT?}, 
  year={2023},
  pages={2445-2451},
  url={https://arxiv.org/abs/2304.09655}}

@InProceedings{vero2025baxbench,
  title = 	 {{B}ax{B}ench: Can {LLM}s Generate Correct and Secure Backends?},
  author =       {Vero, Mark and M\"{u}ndler, Niels and Chibotaru, Victor and Raychev, Veselin and Baader, Maximilian and Jovanovi\'{c}, Nikola and He, Jingxuan and Vechev, Martin},
  booktitle = 	 {Proceedings of the 42nd International Conference on Machine Learning},
  pages = 	 {61344--61390},
  year = 	 {2025},
  volume = 	 {267},
  series = 	 {Proceedings of Machine Learning Research},
  month = 	 {13--19 Jul},
  publisher =    {PMLR},
  url = 	 {https://proceedings.mlr.press/v267/vero25a.html},

}

@article{zhao2025vibecoding,
  title={Is vibe coding safe? Benchmarking vulnerability of agent-generated code in real-world tasks},
  author={Zhao, Songwen and Wang, Danqing and Zhang, Kexun and Luo, Jiaxuan and Li, Zhuo and Li, Lei},
  journal={arXiv preprint arXiv:2512.03262},
  year={2025},
  url={https://arxiv.org/abs/2512.03262}
}

@misc{chen2025securevibebench,
      title={SecureVibeBench: Evaluating Secure Coding Capabilities of Code Agents with Realistic Vulnerability Scenarios}, 
      author={Junkai Chen and Huihui Huang and Yunbo Lyu and Junwen An and Jieke Shi and Chengran Yang and Ting Zhang and Haoye Tian and Yikun Li and Zhenhao Li and Xin Zhou and Xing Hu and David Lo},
      year={2026},
      eprint={2509.22097},
      archivePrefix={arXiv},
      primaryClass={cs.SE},
      url={https://arxiv.org/abs/2509.22097}, 
}

\end{document}